\documentstyle[12pt,citesort,subeqn]{article}
\begin{document}
\newcommand{\vecvar}[1]{\mbox{\boldmath$#1$}}
\def\tsum{\mathop{{\rm T}_{\Sigma}}}
\newcommand{\myref}[1]{(\ref{#1})}
\def\Define{\mathop{\stackrel{\rm def}{=}}}
\def\Hchat{{{\hat H}_{\rm C}}}
\def\Hstilde{{{\tilde H}_{\rm S}}}
\def\Hc{{H_{\rm C}}}
\def\Hs{{H_{\rm S}}}
\def\Lcheck{{\check L}}
\def\Qcheck{{\check Q}}
\def\Lm{{L^{-}}}
\def\Lp{{L^{+}}}
\def\Lmhat{{{\hat L}^{-}}}
\def\Lphat{{{\hat L}^{+}}}
\def\phighat{{{\hat \phi}_{\rm g}}}
\def\varphighat{{{\hat \varphi}_{\rm g}}}
\def\psigtilde{{{\tilde \psi}_{\rm g}}}
\def\dalpha{\alpha^{\dagger}}
\def\Ltilde{{{\tilde L}}}
\def\Qtilde{{{\tilde Q}}}
\def\Lcal{{\tilde {\cal L}}}
\def\ledo{{\stackrel{\rm D}{\leq}}}
\def\Sym{\Bigr|_{\rm Sym}}
\def\bp{b^{+}}
\def\Mae{\mbox{\hspace*{-20pt}}}
\def\Cmae{\mbox{\hspace*{-10pt}}}
\def\Usr{\mbox{\hspace*{20pt}}}
\def\Cusr{\mbox{\hspace*{10pt}}}
\hspace*{\fill}{\sf UT/Wu015, cond-mat/9706133}
\begin{flushleft}

{\large\bf Orthogonal Symmetric Polynomials Associated with

the Calogero Model}

\vspace{10pt}

{\large Hideaki Ujino}

\vspace{10pt}

{\large\it Department of Physics, Graduate School of Science,

University of Tokyo,

Hongo 7--3--1, Bunkyo-ku, Tokyo 113, Japan}

\end{flushleft}\vspace{10pt}

\noindent{\bf Abstract.}
The Calogero model is a one-dimensional quantum integrable system with
inverse-square long-range interactions confined in an external harmonic
well.
It shares the same algebraic structure with the Sutherland model,
which is also
a one-dimensional quantum integrable system with inverse-sine-square
interactions. Inspired by the Rodrigues formula for the Jack polynomials,
which form the orthogonal basis of the Sutherland model,
recently found by Lapointe and Vinet, we construct the Rodrigues
formula for
the Hi-Jack (hidden-Jack) polynomials that form the orthogonal
basis of the Calogero model.
\vspace{20pt}

Exact solutions for the Schr\"odinger equations have provided important
problems in physics and mathematical physics.
Most of us have studied the Laguerre polynomials and the
spherical harmonics in the theory of the hydrogen atom,
and the Hermite polynomials and their Rodrigues formula in the theory
of the quantum harmonic oscillator.
The former is also a good example that shows
the role of conserved operators in quantum mechanics.
The hydrogen atom has
three, independent and mutually commuting conserved operators, namely,
the Hamiltonian, the total angular momentum and its $z$-axis component.
The simultaneous eigenfunctions for the three conserved
operators give the
orthogonal basis of the hydrogen atom.
A classical system with a set of independent and
mutually Poisson commuting (involutive) conserved quantities whose
number of elements is the same as the
degrees of freedom of the system can be integrated by quadrature.
This is guaranteed by the Liouville theorem.
Such a system is called the completely integrable system.
Quantum systems with enough number of such conserved operators are
analogously called quantum integrable systems.
The hydrogen atom is a simple example of the quantum integrable system.

Among the various quantum integrable systems, one-dimensional quantum
many-body systems with inverse-square long-range interactions are now
attracting much interests of theoretical physicists.
Of the various integrable inverse-square-interaction models, the
Calogero model~\cite{Calogero_1} has the longest history.
Its Hamiltonian is
expressed as
\begin{equation}
  \Mae \Hchat = 
  \frac{1}{2}\sum_{j=1}^{N}\bigl(p_{j}^{2}+\omega^{2}x_{j}^{2}\bigr)
  +\frac{1}{2}\sum_{\stackrel{\scriptstyle j,k=1}{j\neq k}}^{N}
  \frac{a^{2}-a}{(x_{j}-x_{k})^{2}},
  \label{eqn:Calogero_model}
\end{equation}
where the constants $N$, $a$ and $\omega$ are the particle number,
the coupling parameter and the strength of the external harmonic well,
respectively.
The momentum operator $p_{j}$ is given by a differential operator,
$p_{j} = -{\rm i}\frac{\partial}{\partial x_{j}}$.
Throughout the paper, we set the Planck constant at unity,
$\hbar=1$.
This model is known to be a quantum integrable system in the sense that
it has sufficient number of independent and mutually commuting conserved
operators \cite{Polychronakos_1,Ujino_1,Ujino_2,Ujino_3,Ujino_4}.
On the other hand, the Sutherland model \cite{Sutherland_1},
which is also a one-dimensional quantum integrable system with
inverse-sine-square interactions,
\begin{equation}
  \Hstilde=\frac{1}{2}\sum_{j=1}^{N}p_{j}^{2}
  +\frac{1}{2}\sum_{\stackrel{\scriptstyle j,k=1}{j\neq k}}^{N}
  \frac{a^{2}-a}{\sin^{2}(x_{j}-x_{k})},
  \label{eqn:Sutherland_model}
\end{equation}
has been thoroughly investigated and its orthogonal basis is known to
consist of the Jack symmetric polynomials
\cite{Jack_1,Stanley_1,Macdonald_1}.
The well investigated properties of the Jack polynomials
made it possible to exactly calculate
correlation functions of the Sutherland model
\cite{Forrester_1,Forrester_2,Lesage_1,Ha_1}.
On the other hand,
the orthogonal basis of the Calogero model was not familiar
with theoretical physicists.
Thus we wanted to identify the orthogonal basis
of the Calogero model that must play an important role
in calculation of
various correlation functions of the Calogero model.

In order to identify the orthogonal basis of the Calogero model,
we have been applying a naive approach which we use in the study of
the hydrogen atom.
We have been investigating the simultaneous eigenfunction
for all the conserved operators of the Calogero model that should
be a deformed multivariable extension of the Hermite polynomial
\cite{Ujino_5}. For the Calogero model,
some ways of algebraic construction of the energy eigenfunctions
are known
\cite{Ujino_1,Ujino_2,Brink_1,Brink_2}. Taking linear combinations
of algebraically constructed energy eigenfunctions
\cite{Ujino_1,Ujino_2},
we tried to make simultaneous eigenfunctions for the conserved operators
\cite{Ujino_5}, though the method was not a practical way to deal
with general cases.
The crucial hints for us are the common algebraic
structure for the Calogero and the Sutherland models
and the Rodrigues formula for the Jack polynomials recently found
by Lapointe and Vinet \cite{Lapointe_1,Lapointe_2}.
The quantum Lax formulation and the Dunkl operator \cite{Dunkl_1}
(exchange operator) formulation showed that these two models 
have the same algebraic 
structure \cite{Polychronakos_1,Ujino_1,Ujino_2,Ujino_3,Ujino_4,%
Bernard_1,Bernard_2,Hikami_1,Hikami_2}.
The fact strongly suggests some similarity in the structures of their
Hilbert spaces.
Because of the common algebraic structure that the Calogero model and the
Sutherland model share, the simultaneous eigenfunctions must be similar
to the Jack polynomials.
Thus we call the unidentified simultaneous eigenfunctions of the Calogero
model Hi-Jack (hidden-Jack) polynomials
\cite{Ujino_6,Ujino_7,Ujino_8}.
We shall extend the method Lapointe and Vinet developed to
construct the Rodrigues formula for the Jack polynomials
\cite{Lapointe_1,Lapointe_2} to the quantum Calogero model and derive
the Rodrigues formula for the Hi-Jack polynomials
\cite{Ujino_6,Ujino_7}.
We shall study some properties of the Hi-Jack polynomials such as
integrality, triangularity and orthogonality.
We shall also study the relationships and similarities
between the Jack polynomials and the Hi-Jack polynomials
\cite{Ujino_7,Ujino_8}.

First, we formulate the Dunkl operator formulation
for the Calogero model.
The ground state wave function is the real Laughlin wave function:
\begin{equation}
  \phighat = \prod_{1\leq j<k\leq N}|x_{j}-x_{k}|^{a}
  \exp\bigl(-\frac{1}{2}\omega\sum_{j=1}^{N}x_{j}^{2}\bigr).
  \label{eqn:Real_Laughlin}
\end{equation}
A short note might be in order.
The phase of the difference product of the above
real Laughlin wave function,
which determines the statistics of the particles,
or in other words, the symmetry of all the eigenfunctions,
can be arbitrary.
We can assign any phase factor to the exchanges of particles.
However,
we must introduce a phase factor to the definition of the Dunkl
operators \cite{Ujino_4}.
To avoid unnecessary complexity, we fix the phase at unity.

The eigenfunction of the Calogero model is fatorized into an
inhomogeneous symmetric polynomial and the ground state wave function.
For convenience of investigations on the inhomogeneous
symmetric polynomials, we introduce a notation.
Any operator with a hat ${\hat {\cal O}}$ is related to
an operator ${\cal O}$ by the similarity transformation using the ground
state wave function $\phighat$,
\begin{subequations}
  \begin{eqnarray}
    {\cal O} & = & \phighat^{-1}{\hat {\cal O}}\phighat,
    \label{eqn:Harmonic_similar_transformation}\\
    {\hat {\cal O}} & = & \phighat{\cal O}\phighat^{-1}.
    \label{eqn:Inverse_harmonic_similar_transformation}
  \end{eqnarray}
  \label{eqn:Harmonic_similar}
\end{subequations}
The above similarity transformation removes the action
on the ground state wave function
\myref{eqn:Real_Laughlin} from any operator
with hat $\hat{\cal O}$.

The Dunkl operators for the model are given by
\begin{subequations}
  \begin{eqnarray}
    \alpha_{l} & = & {\rm i}\bigl(p_{l}
    +{\rm i}a\sum_{\stackrel{\scriptstyle k=1}{k\neq l}}^{N}
    \frac{1}{x_{l}-x_{k}}(K_{lk}-1)\bigr),
    \label{eqn:Annihilation}\\
    \dalpha_{l} & = & -\frac{\rm i}{2\omega}\bigl(p_{l}
    +{\rm i}a\sum_{\stackrel{\scriptstyle k=1}{k\neq l}}^{N}
    \frac{1}{x_{l}-x_{k}}(K_{lk}-1) +2{\rm i}\omega x_{l}\bigr),
    \label{eqn:Creation}\\
    d_{l} & = & \dalpha_{l}\alpha_{l},
    \label{eqn:d_operator}
  \end{eqnarray}
  \label{eqn:Dunkl}
\end{subequations}
where $K_{lk}$ is the coordinate exchange operator.
The operator $K_{lk}$ has the properties
\begin{eqnarray}
  & & K_{lk} = K_{kl}, \;\; (K_{lk})^{2} = 1, \;\;
  K_{lk}^{\dagger}=K_{lk},\;\;
  K_{lk}\cdot 1 = 1,\nonumber\\
  & & K_{lk}A_{l}=A_{k}K_{lk}, \;\;
  K_{lk}A_{j} = A_{j}K_{lk},\;\;
  \mbox{for }j\neq l,k,
  \label{eqn:Exchange_operator}
\end{eqnarray}
where $A_{j}$ is either a partial differential operator 
$\frac{\partial}{\partial x_{j}}$ (or equivalently,
a momentum operator $p_{j}$), a particle coordinate
$x_{j}$ or coordinate exchange operators $K_{jk}$, $k=1,2,\cdots,N$, 
$k\neq j$.
The above properties of the coordinate exchange operator
are also expressed as the action on a multivariable function:
\begin{equation}
  (K_{lk}f)(x_{1},\cdots,x_{l},\cdots,x_{k},\cdots,x_{N})=
  f(x_{1},\cdots,x_{k},\cdots,x_{l},\cdots,x_{N}).
  \label{eqn:Exchange_action}
\end{equation}
Note that the action on the ground state of the above Dunkl
operators has
already been removed by the similarity transformation
\myref{eqn:Harmonic_similar}, and hence they are operators without hat.
Commutation relations among the Dunkl operators and the action of
$\alpha_{l}$ on $1$ are
\begin{subequations}
  \begin{eqnarray}
    & & [\alpha_{l},\alpha_{m}] = 0,\;\;\;
    [\dalpha_{l},\dalpha_{m}] = 0,
    \label{eqn:Commutator_harmonic_Dunkl_0}\\
    & & [\alpha_{l},\dalpha_{m}]
    = \delta_{lm}
    \bigl(1+a\sum_{\stackrel{\scriptstyle k=1}{k\neq l}}^{N}K_{lk}\bigr)
    -a(1-\delta_{lm})K_{lm},\label{eqn:Commutator_harmonic_Dunkl_1}\\
    & & [d_{l},d_{m}] = a(d_{m}-d_{l})K_{lm},
    \label{eqn:Commutator_harmonic_Dunkl_2}\\
    & & \alpha_{l} \cdot 1 = 0.
    \label{eqn:Action_harmonic_Dunkl}
  \end{eqnarray}
  \label{eqn:Needed_algebra_harmonic_Dunkl}
\end{subequations}
We should remark that the above relations do not explicitly depend on
the parameter $\omega$.
As we have mentioned, the phase factor of the difference
product part of the
ground state wave function can be
arbitrary. This phase factor affects the definition of
the Dunkl operators and
coordinate exchange operators
with hat, i.e., 
${\hat \alpha}_{l}$, ${\hat \dalpha}_{l}$, ${\hat d}_{l}$
and ${\hat K_{lk}}$.
We may introduce a phase factor in the defining relations of the
coordinate exchange operators \myref{eqn:Exchange_operator} and the
commutation relations of the Dunkl operators
\myref{eqn:Needed_algebra_harmonic_Dunkl} with hat \cite{Ujino_4}.
This modification is naturally
introduced by the inverse of the similarity transformation
\myref{eqn:Inverse_harmonic_similar_transformation}
of the Dunkl operators without hat \myref{eqn:Dunkl}.

A set of mutually commuting conserved operators 
of the Calogero model $\{I_{n}|n=1,2,\cdots,N\}$ is given by
\begin{equation}
  I_{n} = 
  \sum_{l=1}^{N}(d_{l})^{n}\Bigr|_{\rm Sym},\;
  [I_{n},I_{m}]=0, \; n,m =1,2,\cdots,N,
  \label{eqn:Calogero_conserved_operators_Dunkl}
\end{equation}
where the symbol $\Bigr|_{\rm Sym}$ means that the operand
is restricted to symmetric functions \cite{Ujino_7}.
The Hamiltonian $\Hc$, which is related to the original 
Calogero Hamiltonian
\myref{eqn:Calogero_model} by eq. \myref{eqn:Harmonic_similar},
is equal to $\omega I_{1}+E_{\rm g}$. 
From now on,
we regard the first conserved operator $I_{1}$ as 
the Hamiltonian of the Calogero model.

Next, we formulate the Dunkl operator formulation for 
the Sutherland model. By the change of the variables,
\begin{equation}
  \exp{2{\rm i}x_{j}}=z_{j},\;\;\; j=1,2,\cdots,N,
  \label{eqn:Variable_change}
\end{equation}
the Hamiltonian of the Sutherland model \myref{eqn:Sutherland_model} is
transformed to
\begin{equation}
  \Hstilde = -2\Bigl(\sum_{j=1}^{N}(z_{j}p_{z_{j}})^{2}
  +(a^{2}-a)\sum_{\stackrel{\scriptstyle j,k=1}{j\neq k}}^{N}
  \frac{z_{j}z_{k}}{(z_{j}-z_{k})^{2}}\Bigr),
  \label{eqn:Transformed_Sutherland_model}
\end{equation}
where $p_{z_{j}}=-{\rm i}\frac{\partial}{\partial z_{j}}$.
The ground state wave function and the ground state energy for the above
Hamiltonian \myref{eqn:Transformed_Sutherland_model} are
\begin{eqnarray}
  \psigtilde & = & \prod_{1\leq j<k\leq N}|z_{j}-z_{k}|^{a}
  \prod_{j=1}^{N}z_{j}^{-\frac{1}{2}a(N-1)},
  \label{eqn:Trigono_Jastraw_in_z}\\
  \epsilon_{\rm g} & = & \frac{1}{6}a^{2}N(N-1)(N+1).
  \label{eqn:Sutherland_GE}
\end{eqnarray}
In a similar way to the similarity transformation for the operators
related to the Calogero model, we introduce a similarity transformation
for the operators related to the Sutherland model using the above
ground state wave function \myref{eqn:Trigono_Jastraw_in_z}.
The similarity transformation of the above Hamiltonian yields
\begin{eqnarray}
  \Hs-\epsilon_{\rm g}
  & = & \psigtilde^{-1}(\Hstilde-\epsilon_{\rm g})\psigtilde
  \nonumber\\
  & = & -2\sum_{j=1}^{N}(z_{j}p_{z_{j}})^{2} + {\rm i}a
  \sum_{\stackrel{\scriptstyle j,k=1}{j\neq k}}^{N}
  \frac{z_{j}+z_{k}}{z_{j}-z_{k}}(z_{j}p_{z_{j}}-z_{k}p_{z_{k}}).
  \label{eqn:Projected_Sutherland}
\end{eqnarray}
This operator will be derived from the Dunkl operators for
the Sutherland model.

The Dunkl operators for the Sutherland model,
whose action on the ground state
\myref{eqn:Trigono_Jastraw_in_z} is removed
in a similar way to deal with the Dunkl operators for 
the Calogero model,
are
\begin{subequations}
  \begin{eqnarray}
    \nabla_{l} & = & {\rm i}\bigl(p_{z_{l}}
    +{\rm i}a\sum_{\stackrel{\scriptstyle k=1}{k\neq l}}^{N}
    \frac{1}{z_{l}-z_{k}}(K_{lk}-1)\bigr),
    \label{eqn:Coupled_momentum}\\
    z_{l}, & &
    \label{eqn:coordinate}\\
    D_{l} & = & z_{l}\nabla_{l}.
    \label{eqn:D_operator}
  \end{eqnarray}
  \label{eqn:Sutherland_Dunkl}
\end{subequations}
These Dunkl operators satisfies the following relations,
\begin{subequations}
  \begin{eqnarray}
    & & [\nabla_{l},\nabla_{m}] = 0,\;\;\;
    [z_{l},z_{m}] = 0,
    \label{eqn:Commutator_Sutherland_Dunkl_0}\\
    & & [\nabla_{l},z_{m}] = \delta_{lm}
    \bigl(1+a\sum_{\stackrel{\scriptstyle k=1}{k\neq l}}^{N}K_{lk}\bigr)
    -a(1-\delta_{lm})K_{lm},\label{eqn:Commutator_Dunkl_Sutherland_1}\\
    & & [D_{l},D_{m}] = a(D_{m}-D_{l})K_{lm},
    \label{eqn:Commutator_Dunkl_Sutherland_2}\\
    & & \nabla_{l} \cdot 1 = 0.
    \label{eqn:Action_Dunkl_Sutherland}
  \end{eqnarray}
  \label{eqn:Needed_algebra_Dunkl_Sutherland}
\end{subequations}
which are completely the same as those of Dunkl operators for
the Calogero model \myref{eqn:Needed_algebra_harmonic_Dunkl}.
Commuting conserved operators are also written in a similar way to eq.
\myref{eqn:Calogero_conserved_operators_Dunkl} by the Dunkl operator as
\begin{equation}
  {\cal I}_{n}=\sum_{l=1}^{N}(D_{l})^{n}\Bigr|_{\rm Sym},\;
  [{\cal I}_{n},{\cal I}_{m}]=0,\;n,m=1,2,\cdots,N.
  \label{eqn:Sutherland_conserved_operators_Dunkl}
\end{equation}
The Hamiltonian \myref{eqn:Projected_Sutherland}
corresponds to the second conserved operator,
$H_{\rm S}-\epsilon_{\rm g}=2{\cal I}_{2}$.
Comparing the two Dunkl operator formulations,
we notice the correspondence between the two sets of Dunkl operators:
\begin{equation}
  \alpha_{l} \leftrightarrow \nabla_{l},
  \dalpha_{l} \leftrightarrow z_{l},
  d_{l} \leftrightarrow D_{l}.
  \label{eqn:Correspondence_two_Dunkl}
\end{equation}
This correspondence exhibits the fact that
the Calogero and the Sutherland models share a common algebraic structure.
Moreover, when we identify $z_{l}$ with $x_{l}$, $l=1,\cdots,N$,
the Dunkl
operators for the Calogero model reduce to those for the Sutherland model
in the limit $\omega\rightarrow\infty$.
Thus our theory for the Hi-Jack symmetric polynomials described by
the Dunkl operators for the Calogero model is a one-parameter deformation
of the theory for
the Jack symmetric polynomials written by the Dunkl operators for the
Sutherland model.

Now we introduce the Hi-Jack polynomials.
Symmetric polynomials with $N$ variables are labeled by the Young
tableaux $Y_{N}$.
A Young tableau, $\lambda\in Y_{N}$, is a set of $N$ nonnegative integers
arranged in non-increasing order,
\begin{equation}
  \lambda = (\lambda_{1}\geq\lambda_{2}\geq\cdots\geq\lambda_{N}
  \geq 0 ).
  \label{eqn:Young_tableau}
\end{equation}
We sometimes use a compact notation for the Young tableaux, for example,
$(3^{3},2^{4},1)=(3,3,3,2,2,2,2,1,0,\cdots,0)$. As a basis of the
symmetric polynomials, we often use the monomial symmetric polynomials,
or in other words, symmetrized monomials $m_{\lambda}$ defined by
\begin{equation}
  m_{\lambda}(x_{1},x_{2},\cdots,x_{N})
  = \sum_{
    \stackrel{\scriptstyle \sigma:\;\mbox{\scriptsize distinct}}
    {\mbox{\scriptsize permutation}}}
    (x_{\sigma(1)})^{\lambda_{1}}
    (x_{\sigma(2)})^{\lambda_{2}}\cdots
    (x_{\sigma(N)})^{\lambda_{N}}.
  \label{eqn:symmetrized_monomial}
\end{equation}
Note that the summation over $S_{N}$ is performed so that any monomial
in the summand appears only once.
We denote the Jack polynomial and the Hi-Jack polynomial
that are labeled by a Young tableau $\lambda$ by
$J_{\lambda}(\vecvar{x};1/a)$ and $j_{\lambda}(\vecvar{x};\omega,1/a)$
respectively, where $\vecvar{x}=(x_{1},x_{2},\cdots,x_{N})$.
Following a definition of the Jack symmetric polynomials
$J_{\lambda}(\vecvar{x};1/a)$, 
we define the Hi-Jack symmetric polynomials
$j_{\lambda}(\vecvar{x};\omega,1/a)$ by
\begin{subequations}
  \begin{eqnarray}
    I_{1}j_{\lambda}(\vecvar{x};\omega,1/a) & = &
    \sum_{k=1}^{N}\lambda_{k}j_{\lambda}(\vecvar{x};\omega,1/a)
    \nonumber\\
    & = & E_{1}(\lambda)j_{\lambda}(\vecvar{x};\omega,1/a),
    \label{eqn:Hi-Jack_eigenfunction_1}\\
    I_{2}j_{\lambda}(\vecvar{x};\omega,1/a)
    & = & \sum_{k=1}^{N}
    \bigl(\lambda_{k}^{2}+a(N+1-2k)\lambda_{k}\bigr)
    j_{\lambda}(\vecvar{x};\omega,1/a)\nonumber\\
    & = & E_{2}(\lambda)j_{\lambda}(\vecvar{x};\omega,1/a),
    \label{eqn:Hi-Jack_eigenfunction_2}\\
    j_{\lambda}(\vecvar{x};\omega,1/a)
    & = & \sum_{\stackrel{\scriptstyle \mu\ledo\lambda}
    {\mbox{\scriptsize or }|\mu|<|\lambda|}}
    w_{\lambda\mu}(a,1/2\omega)
    m_{\lambda}(\vecvar{x}),
    \label{eqn:Hi-Jack_triangularity}\\
    w_{\lambda\lambda}(a,\omega) & = & 1,
    \label{eqn:Hi-Jack_normalization}
  \end{eqnarray}
  \label{eqn:Hi-Jack_definition}
\end{subequations}
where $|\lambda|$ is the weight of the Young tableau,
${\displaystyle |\lambda|=\sum_{k=1}^{N}\lambda_{k}}$.
The symbol $\ledo$ is the dominance order among the Young
tableaux \cite{Stanley_1,Macdonald_1}:
\begin{equation}
  \mu\ledo\lambda\Leftrightarrow
  \sum_{k=1}^{N}\mu_{k}=\sum_{k=1}^{N}\lambda_{k}
  \mbox{ and }\sum_{k=1}^{l}\mu_{k}\leq\sum_{k=1}^{l}\lambda_{k}
  \mbox{ for all } l.
  \label{eqn:Dominance_ordering}
\end{equation}
Note that the dominance order is not a total order but a partial
order. Thus sometimes we cannot define the dominance order between
some pair of the Young tableaux, e.g. $(3,1^{3})$ and $(2^{3})$.
The first two formulae of the above definition mean that the Hi-Jack
polynomials are simultaneous eigenfunctions for the first two
conserved operators of the Calogero model. The third formula is called
the triangularity, which means that
only the monomial symmetric polynomials labeled by the
Young tableaux $\mu$ satisfying $\mu\ledo\lambda$ or $|\mu|<|\lambda|$
appear in the expansion of the Hi-Jack polynomials with respect to
the monomial symmetric polynomials. The fourth one specifies the
normalization.

Triangularity of the Hi-Jack polynomials with respect to
the dominance order plays an essential role in the unique
identification of the Hi-Jack polynomials. Since the
eigenvalues for only the first two out of the $N$ commuting
conserved operators are given,
an eigenfunction cannot be uniquely identified by the two
eigenvalues because of remaining
degeneracy. For example, the two Young tableaux $(3,1^{3})$
and $(2^{3})$ give the same first two eigenvalues:
\begin{equation}
  E_{1}(3,1^{3})=E_{1}(2^{3})=6,\;\;
  E_{2}(3,1^{3})=E_{2}(2^{3})=12+6a(N-3).
\end{equation}
However, we cannot define the dominance order between the two Young
tableaux, as we have mentioned before. Generally speaking,
we cannot define the dominance order between any pair of distinct
Young tableaux $\lambda$ and $\mu$ that share the common first two
eigenvalues, 
$E_{1}(\lambda)=E_{1}(\mu)$ and $E_{2}(\lambda)=E_{2}(\mu)$
\cite{Ujino_8}.
So combining the eigenvalues and triangularity,
we can uniquely identify the Hi-Jack polynomials.

In order to write down the Rodrigues formula 
for the Hi-Jack polynomials,
it is convenient to introduce the following operators,
\begin{subequations}
  \begin{eqnarray}
    \dalpha_{J} & = & \prod_{j\in J}\dalpha_{j},
    \label{eqn:Monomial_generator}\\
    d_{m,J} & = & (d_{j_{1}} + ma)(d_{j_{2}} + (m+1)a) 
    \cdots(d_{j_{k}} + (m+k-1)a),
    \label{eqn:Q-det}
  \end{eqnarray}
  \label{eqn:Parts_of_generator}
\end{subequations}
where $J$ is a subset of a set $\{1,2,\cdots,N\}$
whose number of elements $|J|$ is
equal to $k$, $J\subseteq\{1,2,\cdots,N\}$, $|J|=k$. 
From eq. \myref{eqn:Commutator_harmonic_Dunkl_2}, 
we can verify an identity,
\begin{equation}
  (d_{i}+ma)(d_{j}+(m+1)a)\Sym^{\{i,j\}}
  = (d_{j}+ma)(d_{i}+(m+1)a)\Sym^{\{i,j\}},
  \label{eqn:Identity_di_and_dj}
\end{equation}
where $m$ is some integer. The symbol $\Sym^{J}$ where $J$ 
is some set of
integers means that the operands are restricted to the space that
is symmetric with respect to the exchanges of any indices
in the set $J$. 
This identity \myref{eqn:Identity_di_and_dj}
guarantees that the operator $d_{m,J}$ does not depend on the order
of the elements of a set $J$ when it acts on symmetric functions.
The raising operators of the Hi-Jack polynomials are expressed as
\begin{subequations}
  \begin{eqnarray}
    & & b^{+}_{k} =
    \sum_{\stackrel{\scriptstyle J\subseteq \{1,2,\cdots,N\}}{|J|=k}}
    \dalpha_{J}d_{1,J},\;\mbox{for }
    k=1,2,\cdots,N-1,
    \label{eqn:Generator_bulk}\\
    & & b^{+}_{N} =
    \dalpha_{1}\dalpha_{2}\cdots\dalpha_{N}.
    \label{eqn:Generator_edge}
  \end{eqnarray}
  \label{eqn:Generator}
\end{subequations}
Using the raising operators
\myref{eqn:Generator}, we can express the Rodrigues formula
for the Hi-Jack polynomials $j_{\lambda}(\vecvar{x};\omega,1/a)$ as
\begin{equation}
  j_{\lambda}(\vecvar{x};\omega,1/a) =
  C_{\lambda}^{-1}(b_{N}^{+})^{\lambda_{N}}
  (b_{N-1}^{+})^{\lambda_{N-1}-\lambda_{N}}
  \cdots (b_{1}^{+})^{\lambda_{1}-\lambda_{2}}\cdot 1,
  \label{eqn:Hi-Jack_polynomials}
\end{equation}
with the normalization constant $C_{\lambda}$ given by
\begin{equation}
  C_{\lambda}=
  \prod_{k=1}^{N-1}C_{k}(\lambda_{1},\lambda_{2},\cdots,\lambda_{k+1};a),
  \label{eqn:Normalization}
\end{equation}
where
\begin{equation}
  C_{k}(\lambda_{1},\lambda_{2},\cdots,\lambda_{k+1};a)
  = (a)_{\lambda_{k}-\lambda_{k+1}}
  (2a+\lambda_{k-1}-\lambda_{k})_{\lambda_{k}-\lambda_{k+1}}
  \cdots(ka+\lambda_{1}-\lambda_{k})_{\lambda_{k}-\lambda_{k+1}}.
  \label{eqn:Parts_of_normalization}
\end{equation}
In the above expression, the symbol $(\beta)_{n}$ is the
Pochhammer symbol, that is, $(\beta)_{n}=\beta(\beta+1)\cdots(\beta+n-1)$,
$(\beta)_{0}\Define 1$.
We proved that
the symmetric polynomials generated by
the Rodrigues formula \myref{eqn:Hi-Jack_polynomials} satisfy
the definition of
the Hi-Jack symmetric polynomials \myref{eqn:Hi-Jack_definition}
\cite{Ujino_7}.

The first seven Hi-Jack polynomials are given as
follows:
\begin{subequations}
\begin{eqnarray}
  j_{0}(\vecvar{x};\omega,1/a) & = & J_{0}(\vecvar{x};1/a)
  =m_{0}(\vecvar{x})=1\\
  j_{1}(\vecvar{x};\omega,1/a) & = & J_{1}(\vecvar{x};1/a)
  =m_{1}(\vecvar{x})\\
  j_{1^{2}}(\vecvar{x};\omega,1/a) & = & J_{1^{2}}(\vecvar{x};1/a)
  +\frac{a}{2\omega}\frac{N(N-1)}{2}J_{0}(\vecvar{x};1/a)\nonumber\\
  & = & m_{1^{2}}(\vecvar{x})
  +\frac{a}{2\omega}\frac{N(N-1)}{2}m_{0}(\vecvar{x})\\
  (a+1)j_{2}(\vecvar{x};\omega,1/a) & = & (a+1)J_{2}(\vecvar{x};1/a)
  -\frac{1}{2\omega}N(Na+1)J_{0}(\vecvar{x};1/a)\nonumber\\
  & = & (a+1)m_{2}(\vecvar{x})+2am_{1^{2}}(\vecvar{x})
  -\frac{1}{2\omega}N(Na+1)m_{0}(\vecvar{x})\\
  j_{1^{3}}(\vecvar{x};\omega,1/a) & = & J_{1^{3}}(\vecvar{x};1/a)
  +\frac{1}{2\omega}a\frac{(N-1)(N-2)}{2}J_{1}(\vecvar{x};1/a)
  \nonumber\\
  & = &  m_{1^{3}}(\vecvar{x})
  +\frac{1}{2\omega}a\frac{(N-1)(N-2)}{2}m_{1}(\vecvar{x})\\
  (2a+1)j_{2,1}(\vecvar{x};\omega,1/a) & = &
  (2a+1)J_{2,1}(\vecvar{x};1/a)\nonumber\\
  & & -\frac{1}{2\omega}(1-a)(N-1)(Na+1)
  J_{1}(\vecvar{x};1/a)\nonumber\\
  & = & (2a+1)m_{2,1}(\vecvar{x})+6am_{1^{3}}(\vecvar{x})\nonumber\\
  & & -\frac{1}{2\omega}(1-a)(N-1)(Na+1)m_{1}(\vecvar{x})\\
  (a^{2}+3a+2)j_{3}(\vecvar{x};\omega,1/a) & = &
  (a^{2}+3a+2)J_{3}(\vecvar{x};1/a)
  -\frac{3}{2\omega}(a^{2}N^{2}+3aN+2)J_{1}(\vecvar{x};1/a)
  \nonumber\\
  & = & (a^{2}+3a+2)m_{3}(\vecvar{x})+3a(a+1)m_{2,1}(\vecvar{x})\nonumber\\
  & & +6a^{2}m_{1^{3}}(\vecvar{x})
  -\frac{3}{2\omega}(a^{2}N^{2}+3aN+2)m_{1}(\vecvar{x})
\end{eqnarray}
\end{subequations}
The explicit forms show the fact that the Hi-Jack polynomial is
a one-parameter deformation of the Jack polynomial,
\begin{equation}
  j_{\lambda}(\vecvar{x};\omega=\infty,1/a)=J_{\lambda}(\vecvar{x};1/a).
  \label{eqn:Hi-Jack_to_Jack}
\end{equation}
This has been clarified
in the discussion of the common algebraic structure of the Calogero and
the Sutherland models. Note that the top weight monomial symmetric
polynomials in the expansion of the Hi-Jack polynomials
$j_{\lambda}(\vecvar{x};\omega,1/a)$ form the Jack polynomials
of the same Young tableau $J_{\lambda}(\vecvar{x};1/a)$. This is
because increasing the order of $1/2\omega$ by one causes
decreasing of the weight by two in the expansion
\myref{eqn:Hi-Jack_triangularity}.
We can also observe the triangularity.
Besides the above properties, we notice that all the expansion 
coefficients
are polynomials of $a$ and $1/2\omega$ with integer coefficients.
This observation generally holds.
The expansion coefficients multiplied by the normalization constant
of the Hi-Jack polynomials
$C_{\lambda}w_{\lambda\mu}(a,1/2\omega)$ are
polynomials of $a$ and $1/2\omega$ with integer coefficients,
which is called integrality.
This property is analogous to that stated by
the Macdonald-Stanley conjecture for the Jack
polynomials \cite{Stanley_1,Macdonald_1,Lapointe_2}.

While the Hi-Jack polynomial is a one-parameter deformation
of the Jack polynomial, we can get the Hi-Jack polynomial
from the Jack polynomial by the following formula:
\begin{equation}
  J_{\lambda}(\dalpha_{1},\dalpha_{2},\cdots,\dalpha_{N};1/a)\cdot 1
  = j_{\lambda}(\vecvar{x};\omega,1/a).
  \label{eqn:Jack_to_Hi-Jack}
\end{equation}
Equation \myref{eqn:Jack_to_Hi-Jack} gives another relationship 
between the Jack polynomials and the Hi-Jack polynomials.

We have introduced the Hi-Jack polynomials as the simultaneous
eigenfunctions for the first two commuting conserved operators
with the triangularity.
As we shall see shortly,
they are non-degenerate simultaneous eigenfunctions
for all the commuting conserved operators of the Calogero model.
From a calculation of the action of $d_{l}$ operator on
a symmetrized monomial of $\dalpha_{k}$'s, $m_{\lambda}(\dalpha_{1},
\cdots,\dalpha_{N})$, we can prove the following expression:
\begin{equation}
  I_{n} j_{\lambda}(\vecvar{x};\omega,1/a)=
  \sum_{\stackrel{\scriptstyle \mu\ledo\lambda}
  {\mbox{\scriptsize or }|\mu|<|\lambda|}}
  w^{\prime}_{\lambda,\mu}(a,1/2\omega)m_{\mu}(\vecvar{x}).
  \label{eqn:Keep_triangularity}
\end{equation}
This means that operation of the conserved operators on the Hi-Jack
polynomials keeps their triangularity. Since the $n$-th conserved
operator commutes with the first and second conserved operators,
$[I_{1},I_{n}]=[I_{2},I_{n}]=0$, we can easily verify,
\begin{subequations}
  \begin{eqnarray}
    & & I_{1}I_{n}j_{\lambda}(\vecvar{x};\omega,1/a)=
    E_{1}(\lambda)I_{n}j_{\lambda}(\vecvar{x};\omega,1/a),
    \label{eqn:verification1}\\
    & & I_{2}I_{n}j_{\lambda}(\vecvar{x};\omega,1/a)=
    E_{2}(\lambda)I_{n}j_{\lambda}(\vecvar{x};\omega,1/a).
    \label{eqn:verification2}
  \end{eqnarray}
  \label{eqn:verification}
\end{subequations}
Equations \myref{eqn:verification1}, \myref{eqn:verification2}
and \myref{eqn:Keep_triangularity} for $I_{n}j_{\lambda}$
are respectively the same as eqs. \myref{eqn:Hi-Jack_eigenfunction_1},
\myref{eqn:Hi-Jack_eigenfunction_2} and
\myref{eqn:Hi-Jack_triangularity} for the Hi-Jack polynomial
$j_{\lambda}$, which means $I_{n}j_{\lambda}$ satisfies
the definition of
the Hi-Jack polynomial \myref{eqn:Hi-Jack_definition} except for
normalization.
Our definition of the Hi-Jack polynomials \myref{eqn:%
Hi-Jack_definition} uniquely specifies the Hi-Jack polynomial.
So we conclude
that $I_{n}j_{\lambda}$ must coincide with $j_{\lambda}$ up to
normalization.
Thus we confirm that the Hi-Jack polynomials $j_{\lambda}$
simultaneously diagonalize all the commuting conserved operators
$I_{n}$, $n=1,\cdots,N$. The eigenvalues of the conserved operators,
\begin{equation}
  I_{n}j_{\lambda}(\vecvar{x};\omega,1/a)=E_{n}(\lambda)
  j_{\lambda}(\vecvar{x};\omega,1/a),
  \label{eqn:n_eigenvalue}
\end{equation}
are generally polynomials of the coupling parameter $a$:
\begin{equation}
  E_{n}(a)=e_{n}^{(0)}(\lambda)+e_{n}^{(1)}(\lambda)a+\cdots.
  \label{eqn:eigenvalue_expansion}
\end{equation}
It is easy to get the constant term $e_{n}^{(0)}(\lambda)$
because the
term corresponds to the $n$-th eigenvalue for
$N$ free bosons confined in an external harmonic well:
\begin{equation}
  e_{n}^{(0)}(\lambda)=\sum_{k=1}^{N}(\lambda_{k})^{n}.
  \label{eqn:eigenvalue_constant}
\end{equation}
It is clear that there is no degeneracy in the constant terms of the
eigenvalues $\{e_{n}^{(0)}(\lambda)|n=1,\cdots,N\}$.
Since the conserved operators $I_{n}$ are hermitian operators
concerning the inner product,
\begin{eqnarray}
  \langle j_{\lambda},j_{\mu} \rangle & = & \int_{-\infty}^{\infty}
  \prod_{k=1}^{N}{\rm d}x_{k}|\phighat|^{2}j_{\lambda}j_{\mu}
  \nonumber\\
  & \propto & \delta_{\lambda,\mu},
  \label{eqn:inner_product}
\end{eqnarray}
the Hi-Jack polynomials are the orthogonal symmetric polynomials with
respect to the above inner product.
From the explicit form of the weight function,
\begin{equation}
  |\phighat|^{2}=\prod_{1\leq j<k\leq N}|x_{j}-x_{k}|^{2a}
  \exp\Bigl(-\omega\sum_{l=1}^{N}x_{l}^{2}\Bigr),
  \label{eqn:weight_function}
\end{equation}
we conclude that the Hi-Jack polynomial is a multivariable
generalization of the Hermite polynomial.

Let us summarize our study and comment on some future problems.

Motivated by the success of the Jack polynomials that form the
orthogonal basis of the Sutherland model in the calculations
of correlation functions of the Sutherland model
\cite{Forrester_1,Forrester_2,Lesage_1,Ha_1}, we have tried to
identify the orthogonal basis of the Calogero model
\cite{Ujino_5,Ujino_6,Ujino_7,Ujino_8}.
Based on the fact that
the Calogero model has a set of mutually commuting conserved
operators \cite{Polychronakos_1,Ujino_1,Ujino_2,Ujino_3,Ujino_4},
we have tried to construct the simultaneous eigenfunctions for all the
conserved operators of the Calogero model that must form the orthogonal
basis of the model.
Since the Calogero and the Sutherland models share the common algebraic
structure, it is natural to
introduce the Hi-Jack symmetric polynomials in a similar
way to a definition of the Jack polynomials \cite{Ujino_6,Ujino_7}.
Through their Rodrigues formula that is an extension of the Rodrigues
formula for the Jack symmetric polynomials discovered 
by Lapointe and Vinet
\cite{Lapointe_1,Lapointe_2}, we have clarified that
the Hi-Jack polynomials are the desired simultaneous eigenfunctions
of the Calogero model and hence form the orthogonal
basis of the Calogero model \cite{Ujino_7,Ujino_8}.
Some properties of the Hi-Jack polynomials such as
integrality, triangularity and relationships with the Jack polynomials
are also observed.

From the explicit form of the weight function of the inner product,
we have concluded that the Hi-Jack polynomial is a multivariable
generalization of the Hermite polynomial. According to recent preprints
\cite{Baker_1,Baker_2,Diejen_2},
the generalized Hermite
polynomials were also introduced by Lasselle
\cite{Lasselle_1} and by
Macdonald in an unpublished manuscript. They defined
the generalized Hermite polynomials as orthogonal symmetric polynomials
with respect to the inner product \myref{eqn:inner_product}.
On the other hand, our definition specifies the Hi-Jack polynomials,
or in other words,
the generalized Hermite polynomials,
as the simultaneous eigenfunctions for
the commuting conserved operators of the Calogero model, which are
natural objects for physicists' interest.

Some progresses related to our results were reported recently
\cite{Baker_1,Baker_2,Diejen_2}.
Multivariable generalizations of the classical Laguerre and Jacobi
polynomials, which form the orthogonal basis of Calogero models
associated with root lattices other than $A_{N-1}$
\cite{Olshanetsky_1}, were studied.
Their non-symmetric extensions, which describe the spin or
multi-component generalizations of Calogero models, were also
reported. Further investigations on these
orthogonal polynomials must be important for the study of
Green functions and correlation functions of Calogero models,
which was done for the Sutherland model with the help of
the properties of
the Jack polynomials \cite{Forrester_1,Forrester_2,Lesage_1,Ha_1}.
The quantum Lax formulation and the Dunkl operator formulation
for the Calogero and Sutherland models revealed their $W$-symmetry
and the Yangian symmetry structures
\cite{Ujino_2,Ujino_3,Bernard_1,Hikami_2,Hikami_1,Bernard_2}.
It is interesting to study the orthogonal polynomials 
from the viewpoint
of representation theory of such symmetries.
As examples of such studies, we should note that
the Jack polynomials were identified with the singular vectors of
the Virasoro and $W_{N}$-algebras \cite{Mimachi_1,Awata_1}.
The Macdonald polynomials \cite{Macdonald_1} and their generalizations,
which are $q$-deformations of the orthogonal symmetric polynomials,
are also interesting topics.
The Macdonald polynomials are associated with the discretization, or
in other words, the relativistic generalization of the Sutherland model
\cite{Ruijsenaars_1}. The Rodrigues formula for the Macdonald
polynomials was given \cite{Noumi_1,Noumi_2,Lapointe_3,Lapointe_4,%
Lapointe_5}.
Further studies on the continuous Hahn polynomials \cite{Diejen_3}
which are associated with the relativistic or discretized Calogero
model \cite{Diejen_1} and more generalized $q$-deformed
orthogonal polynomials such as $BC_{N}$-Askey-Wilson polynomials
\cite{Koornwinder_1}
related to the discretized Calogero model associated with
root lattices of $BC_{N}$-type \cite{Diejen_4} are interesting.
We expect to report
some progresses in these directions in the near future.

\section*{Acknowledgements}\hspace*{\parindent}
I would like to express my sincere gratitude to the organizers,
Professors M.-L.~Ge, Y.~Saint-Aubin and L.~Vinet,
for giving me a chance to participate in this meeting and to all
the staffs for their hospitality at Nankai University.
I am also grateful to my supervisor, Professor~Miki~Wadati,
for his continuous encouragement and collaboration.
I appreciate Research Fellowships of the Japan Society for
the Promotion of Science for Young Scientists.


\begin{thebibliography}{99}
\bibitem{Calogero_1} F. Calogero: J. Math. Phys. {\bf 12} (1971) 419.

\bibitem{Polychronakos_1} A. P. Polychronakos:
Phys. Rev. Lett. {\bf 69} (1992) 703.

\bibitem{Ujino_1} H. Ujino and M. Wadati: Chaos, Solitons and Fractals
{\bf 5} (1995) 109.

\bibitem{Ujino_2} H. Ujino and M. Wadati: J. Phys. Soc. Jpn. {\bf 63}
(1994) 3585.

\bibitem{Ujino_3} H. Ujino and M. Wadati: J. Phys. Soc. Jpn. {\bf 64}
(1995) 39.

\bibitem{Ujino_4} H. Ujino and M. Wadati: J. Phys. Soc. Jpn. {\bf 64}
(1995) 4121.

\bibitem{Sutherland_1} B. Sutherland: J. Math. Phys. {\bf 12} (1971) 246.

\bibitem{Jack_1} H. Jack:
Proc. R. Soc. Edinburgh (A) {\bf 69} (1970) 1.

\bibitem{Stanley_1} R. P. Stanley: Adv. Math. {\bf 77} (1988) 76.

\bibitem{Macdonald_1} I. G. Macdonald: {\it Symmetric Functions and Hall
Polynomials} (Clarendon Press, Oxford, 1995) 2nd ed.

\bibitem{Forrester_1} P. J. Forrester: Nucl. Phys. B {\bf 388} (1992) 671.

\bibitem{Forrester_2} P. J. Forrester: Nucl. Phys. B {\bf 416} (1994) 377.

\bibitem{Lesage_1}F. Lesage, V. Pasquier and D. Serban: Nucl. Phys. B
{\bf 435} (1995) 585.

\bibitem{Ha_1} Z. N. C. Ha: Nucl. Phys. B {\bf 435} (1995) 604.

\bibitem{Ujino_5} H. Ujino and M. Wadati: J. Phys. Soc. Jpn. {\bf 64}
(1995) 2703.

\bibitem{Brink_1} L. Brink, T. H. Hansson and M. A. Vasiliev:
Phys. Lett. B {\bf 286} (1992) 109.

\bibitem{Brink_2} L. Brink, T. H. Hansson, S. Konstein and M. A. Vasiliev:
Nucl. Phys. B {\bf 401} (1993) 591.

\bibitem{Lapointe_1} L. Lapointe and L. Vinet: Commun. Math. Phys.
{\bf 178} (1996) 425.

\bibitem{Lapointe_2} L. Lapointe and L. Vinet: IMRN {\bf 9} (1995) 419.

\bibitem{Dunkl_1} C. F. Dunkl: Trans. Amer. Math. Soc. {\bf 311} (1989) 167.

\bibitem{Bernard_1} D. Bernard, M. Gaudin, F. D. M. Haldane and V. Pasquier:
J. Phys. A: Math. Gen. {\bf 26} (1993) 5219.

\bibitem{Bernard_2} D. Bernard, K. Hikami and M. Wadati:
{\it New Developments of Integrable Systems and Long-Ranged Interaction
Models}, ed. M.-L. Ge and Y.-S. Wu (World Scientific Singapore, 1995) p.1.

\bibitem{Hikami_1} K. Hikami and M. Wadati: J. Phys. Soc. Jpn. {\bf 62}
(1993) 4203.

\bibitem{Hikami_2} K. Hikami: J. Phys. A: Math. Gen. {\bf 28} (1995) L131.

\bibitem{Ujino_6} H. Ujino and M. Wadati: J. Phys. Soc. Jpn. {\bf 65}
(1996) 653.

\bibitem{Ujino_7} H. Ujino and M. Wadati: J. Phys. Soc. Jpn.
{\bf 65} (1996) 2423.

\bibitem{Ujino_8} H. Ujino and M. Wadati:
J. Phys. Soc. Jpn. {\bf 66} (1997) 345.

\bibitem{Baker_1} T. H. Baker and P. J. Forrester: preprint,
solv-int/9608004.

\bibitem{Baker_2} T. H. Baker and P. J. Forrester: preprint,
solv-int/9609010.

\bibitem{Diejen_2} J. F. van Diejen: preprint, q-alg/9609032.

\bibitem{Lasselle_1} M. Lasselle:
C. R. Acad. Sci. Paris. t. S\'{e}ries I {\bf 313} (1991) 579.

\bibitem{Olshanetsky_1} M. A. Olshanetsky and A. M. Perelomov:
Phys. Rep. {\bf 94} (1983) 313.

\bibitem{Mimachi_1} K. Mimachi and Y. Yamada: Commun. Math. Phys.
{\bf 174} (1995) 447.

\bibitem{Awata_1} H. Awata, Y. Matsuo, S. Odake and J. Shiraishi:
Nucl. Phys. B {\bf 449} (1995) 347.

\bibitem{Ruijsenaars_1} S. N. M. Ruijsenaars: Commun. Math. Phys. {\bf 110}
(1987) 191.

\bibitem{Noumi_1} A. N. Kirillov and M. Noumi: preprint, q-alg/9605004.

\bibitem{Noumi_2} A. N. Kirillov and M. Noumi: preprint, q-alg/9605005.

\bibitem{Lapointe_3} L. Lapointe and L. Vinet: preprint, q-alg/9607024.

\bibitem{Lapointe_4} L. Lapointe and L. Vinet: preprint, q-alg/9607025.

\bibitem{Lapointe_5} L. Lapointe and L. Vinet: preprint, q-alg/9607026.

\bibitem{Diejen_3} J. F. van Diejen: J. Phys. A: Math. Gen. {\bf 28}
(1995) L369.

\bibitem{Diejen_1} J. F. van Diejen: J. Math. Phys. {\bf 36} (1995) 1299.

\bibitem{Koornwinder_1} T. H. Koornwinder: Contemporary Math. {\bf 138}
(1992) 189.

\bibitem{Diejen_4} J. F. van Diejen: Composite Math. {\bf 95} (1995) 183.

\end{thebibliography}
\end{document}